\documentclass[a4paper,fleqn]{cas-dc}

\usepackage[numbers]{natbib}

\def\tsc#1{\csdef{#1}{\textsc{\lowercase{#1}}\xspace}}
\tsc{SNS}

\begin{document}
\let\WriteBookmarks\relax
\def\floatpagepagefraction{1}
\def\textpagefraction{.001}
\shorttitle{OWL}
\shortauthors{S. Neves Silva et~al.}

\title [mode = title]{AutOmatic floW planning for fetaL MRI (OWL)}                      

\tnotetext[1]{This work was supported by the Wellcome Trust, Sir Henry Wellcome Fellowship [201374/Z/16/Z], the UKRI FLF [MR/T018119/1], DFG Heisenberg [502024488], the NIHR Advanced Fellowship [NIHR3016640], the MRC grants [MR/W019469/1] and [MR/X010007/1], and the Wellcome/EPSRC Centre [WT203148/Z/16/Z].}


\author[1,2]{Sara Neves Silva}[type=editor,
      orcid=orcid.org/0009-0009-7520-081X,
      twitter=saranevessilva]
\author[2]{Tomas Woodgate}
\author[1,3]{Sarah McElroy}
\author[1,2]{Michela Cleri}
\author[1,2]{Kamilah St Clair}
\author[1,2]{Jordina Aviles Verdera}
\author[1,2]{Kelly Payette}
\author[1,2]{Alena Uus}
\author[2,4]{Lisa Story}
\author[2]{David Lloyd}
\author[1,2]{Mary A Rutherford}
\author[1,2]{Joseph V Hajnal}
\author[2]{Kuberan Pushparajah}
\author[1,2,5]{Jana Hutter}

\affiliation[1]{organization={Research Department for Early Life Imaging, School of Biomedical Engineering \& Imaging Sciences},
                addressline={King’s College London}, 
                city={London}, 
                country={UK}}

\affiliation[2]{organization={Research Department for Medical Engineering, School of Biomedical Engineering \& Imaging Sciences},
                addressline={King’s College London}, 
                city={London}, 
                country={UK}}

\affiliation[3]{organization={MR Research Collaborations, Siemens Healthcare Limited},
                addressline={Camberley}, 
                country={United Kingdom}}

\affiliation[4]{organization={Department of Women \& Children's Health},
                addressline={King's College London}, 
                city={London}, 
                country={UK}}

\affiliation[5]{organization={Smart Imaging Lab, Radiological Institute},
                addressline={University Hospital Erlangen}, 
                city={Erlangen}, 
                country={Germany}}

\cortext[cor1]{Sara Neves Silva}

\begin{abstract}
\noindent \textbf{Purpose:} Widening access to quantitative fetal flow imaging through fully automated real-time planning of 2D phase-contrast flow imaging of the major fetal vessels (OWL). \textbf{Methods:} Two subsequent deep learning networks, one localizing the fetal chest and one identifying a set of landmarks on a coronal whole-uterus balanced steady-state free precession scan, were trained on 167 and 71 fetal datasets across field strengths, acquisition parameters, and gestational ages and implemented in a real-time setup. Next, a phase-contrast sequence was modified to use the identified landmarks for planning. The OWL pipeline was evaluated retrospectively in 10 datasets and prospectively in 7 fetal subjects (gestational ages between 36+3 and 39+3 weeks). The prospective cases were additionally manually planned to enable direct comparison both qualitatively, by scoring the planning quality, and quantitatively, by comparing the indexed flow measurements. \textbf{Results:} OWL enabled real-time fully automatic planning of the 2D phase-contrast scans in all but one of the prospective participants. The fetal body localization achieved an overall Dice score of 0.94$\pm$0.05 and the cardiac landmark detection accuracy was 5.77±2.91 mm for the descending aorta, 4.32$\pm$2.44 mm for the spine, and 4.94$\pm$3.82 mm for the umbilical vein. For the prospective cases, overall planning quality was 2.73/4 for the automated scans, compared to 3.0/4 for manual planning, and the flow quantitative evaluation showed a mean difference of -1.8\% (range -14.2\% to 14.9\%) by comparing the indexed flow measurements obtained from gated automatic and manual acquisitions. \textbf{Conclusions:} Real-time automated planning of 2D phase-contrast MRI was effectively accomplished for 2 major vessels of the fetal vasculature. While demonstrated here on 0.55T, the achieved method has wider implications, and training across multiple field strengths enables generalization. OWL thereby presents an important step towards extending access to this modality beyond specialised centres.

\end{abstract}


\begin{highlights}
\item Fully automatic real-time planning of 2D Phase-Contrast MRI scans in two of the major fetal vessels - the descending aorta and the umbilical vein - based on a whole-uterus coronal bSSFP scan and using deep learning networks for localization of the fetal chest and identification of fetal cardiac landmarks
\item Quantitative and qualitative evaluation of all parts of the proposed OWL method, with the assessment of the fetal body localization and landmark detection tasks, evaluation of the planning quality, and comparison of the indexed flow measurements obtained from the automatic and manual acquisitions
\item Generalization of the method to other magnetic field strengths, ensuring OWL can be extended to various research and clinical settings beyond specialized centres
\end{highlights}


\maketitle

\section{Introduction}

\noindent In antenatal diagnosis, morphological imaging is typically complemented by quantitative blood flow assessments to provide a comprehensive picture of fetal health. Doppler-based assessments of the major vessels in the heart, umbilical vein \cite{Ferrazzi2000}, uterine artery \cite{Browne2015} and mid-cerebral artery \cite{Rurak2011} are well-established clinical markers employed during antenatal screening. Furthermore, assessing the fetal cardiovascular anatomy and physiology is at the centre of prenatal diagnosis of congenital heart disease (CHD). CHD is the most common fetal anomaly, affecting about 1\% of all live births, often requires surgical and therapeutic interventions in early life and can be associated with subsequent cognitive and motor impairments \cite{Morton2017}.

\noindent Developments in fetal echocardiography have enabled in-utero detection of a wide variety of CHD subtypes, informing both antenatal counselling and postnatal management and in some circumstances facilitating in-utero surgical interventions \cite{Qasim2024}. Such advances have significantly reduced fetal mortality and improved perinatal outcomes in CHD cases. Prenatal disruption of the circulation can lead to effects on other systems, particularly the developing fetal brain \cite{McQuillen2010}. Assessment of Doppler ultrasound (US) waveforms plays a crucial role in prenatal diagnosis, able to identify fetal hypoxia by discerning fetal circulatory adaptations to placental insufficiency \cite{Browne2015} and to inform on flow velocities through the major arteries. However, significant limitations of Doppler US exist when quantifying flow. The typical strategy is to take the cross-sectional area of a vessel and multiply it by the velocity of blood in the centre of the vessel. This assumes a circular vessel, a uniform velocity throughout the cross-section of the vessel, and a constant cross-sectional area throughout the cardiac cycle. Furthermore, it is limited by restrictions in terms of maternal habitus, fetal position and visibility in late gestation. 

\noindent As a result of these limitations, phase-contrast MR (PC-MR) is the diagnostic reference standard for the assessment of hemodynamics and blood flow quantification in postnatal life \cite{Nayak2015}, which can overcome some of the above-mentioned limitations of US. Although MRI is a standard approach in the prenatal evaluation of the fetal central nervous system and increasingly for thoracic or abdominal pathologies \cite{Rodriguez2012,Furey2016,Gat2016}, a comprehensive assessment of the fetal cardiovascular system still lags behind, mainly due to technical limitations. Assessment of the fetal cardiovascular system requires dynamic imaging to resolve cardiac motion and blood flow. To avoid artefacts evoked by cardiac wall motion, synchronization of image acquisition and the cardiac cycle is mandatory. However, the intrauterine position of the fetus does not allow the application of conventional electrocardiogram gating. This has prevented routine application of cardiovascular MRI during the fetal period. Retrospective cardiac gating \cite{jansz2010metric} and MR-compatible Doppler US \cite{Sousa2019} have been around for over a decade, however, while they have shown promise, the required extensive processing and cost of purchase respectively have slowed wider usage.

\noindent Furthermore, a key constraint remains the required accurate planning of the acquisition planes. Accuracy is key to avoiding invalid results due to partial voluming \cite{Jiang2015}. Fetal imaging is typically performed at 1.5T or 3T, however, a recent development for fetal MRI is the reemergence \cite{gowland_vivo_1998} of 0.55T fetal MRI, offering benefits which mitigate common issues: increased field homogeneity reduces distortion artefacts and thus foregoes the need for specialist shimming tools, a larger bore size accommodates pregnant women with higher body mass index (BMI) and in later gestation, decreased RF heating enables more efficient scans and the promise of reduced installation and maintenance cost increases its accessibility. Additionally, the longer T2* value allows longer read-outs and better sampling of the TE dimension e.g., in T2* relaxometry.  However, adaptions are required to compensate for the inherently lower signal-to-noise ratio (SNR). This is typically achieved by reducing the in-plane resolution and/or acquiring thicker slices, putting further emphasis on accurate planning. Recent studies have demonstrated its potential for fetal MRI \cite{aviles2023reliability,ponrartana2023low,payette2023automated}. Part of its appeal is the promise to widen access to fetal MRI beyond specialist centres. This, however, critically relies on also addressing another current bottleneck - the need for specialist resources and skills to conduct fetal MR examinations. One prime example is thereby the accurate planning of fetal flow scans.

\noindent Recent advances leveraged real-time AI frameworks to address planning constraints in adult cardiac MRI. For example, a convolutional neural network (CNN) was developed to label landmarks on cardiac MR images \cite{Xue2021}, achieving a detection rate of 96.6\% to 99.8\%, and producing measurements of cardiac structures comparable to those performed by experts. The workflow was integrated into MRI systems via Gadgetron InlineAI \cite{Xue2019}, with an inference time under 1 second. \cite{Frick2011} introduced a fully automated, operator-independent method for planning standard cardiac geometries and assessing functional parameters in adult patients with various cardiovascular pathologies. Using a model-based segmentation approach for anatomy recognition, an overall success rate of 94\% was achieved in the automatic planning of standard cine views and demonstrated high agreement in cardiac functional parameters. Similarly, fetal head landmark detection was developed for the automatic planning of anatomical fetal brain MRI scans \cite{Neves2024} at 0.55T. Closely related, in fetal cardiac MRI, a study on automatic multi-class fetal vessel segmentation using a combination of deep learning label propagation from multi-class labelled condition-specific atlases and 3D Attention U-Net segmentation was presented \cite{ramirez2022automated}. This approach successfully segmented 12 fetal cardiac vessels and detected three distinct aortic arch anomalies, achieving a 100\% accuracy rate in vessel detection. 

\noindent In this current study, autOmatic floW planning for fetaL MRI (OWL) is proposed, combining a deep learning-based, fast cardiac landmark detection and subsequent automated calculation of the optimal planes for the descending aorta (DAo) and umbilical vein (UV) to achieve fully automatic planning of the acquisition of PC flow MRI scans in two major vessels. The technique was implemented, tested, and evaluated on 10 retrospective and 7 prospective fetal cardiac MR scans. While applied here to 0.55T datasets, the method was trained cross-field strength and is not limited to low-field MRI.

\section*{Theory}
\subsubsection*{Manual planning of the 2D PC vessel acquisitions}
\noindent 2D PC flow assessment requires the positioning of an acquisition slice cross-sectional to the vessel of interest. The manual planning of fetal 2D PC sequences has been previously described \cite{prsa2014}. To inform manual planning of the sequences, balanced steady-state free precession (bSSFP) images of the fetal body are first acquired in the axial, coronal, and sagittal orientations. These are chosen due to a short acquisition time and tissue contrast that enables visualisation of the fetal anatomy. 

\noindent In fetal life, the DAo carries a mixture of oxygenated and de-oxygenated blood to the lower fetal body and, via the umbilical arteries, to the placenta. The anatomy of this vessel is consistent, typically descending inferiorly and coursing anterior and leftward of the spine within the thorax. Flow can thus be measured with a slice in the axial orientation relative to the fetal thorax. This is positioned inferior to the heart with care taken to ensure the slice is superior to the diaphragm, before the branching of the abdominal aorta.

\noindent The UV carries oxygenated blood from the placenta to the fetal abdomen via the umbilical cord, part of this is then preferentially directed to the left atrium via the ductus venosus, with the remainder entering the portal venous system. Whilst PC flow assessment can be performed at any point along its course, the intrahepatic portion is chosen in keeping with that described by \cite{prsa2014}. When manually planning PC acquisitions within this portion, care is taken to position the slice anterior to the portal venous branches, after which the flow will decrease. 


\begin{figure*}[t!]
    \centering
    \includegraphics[width=.65\textwidth]{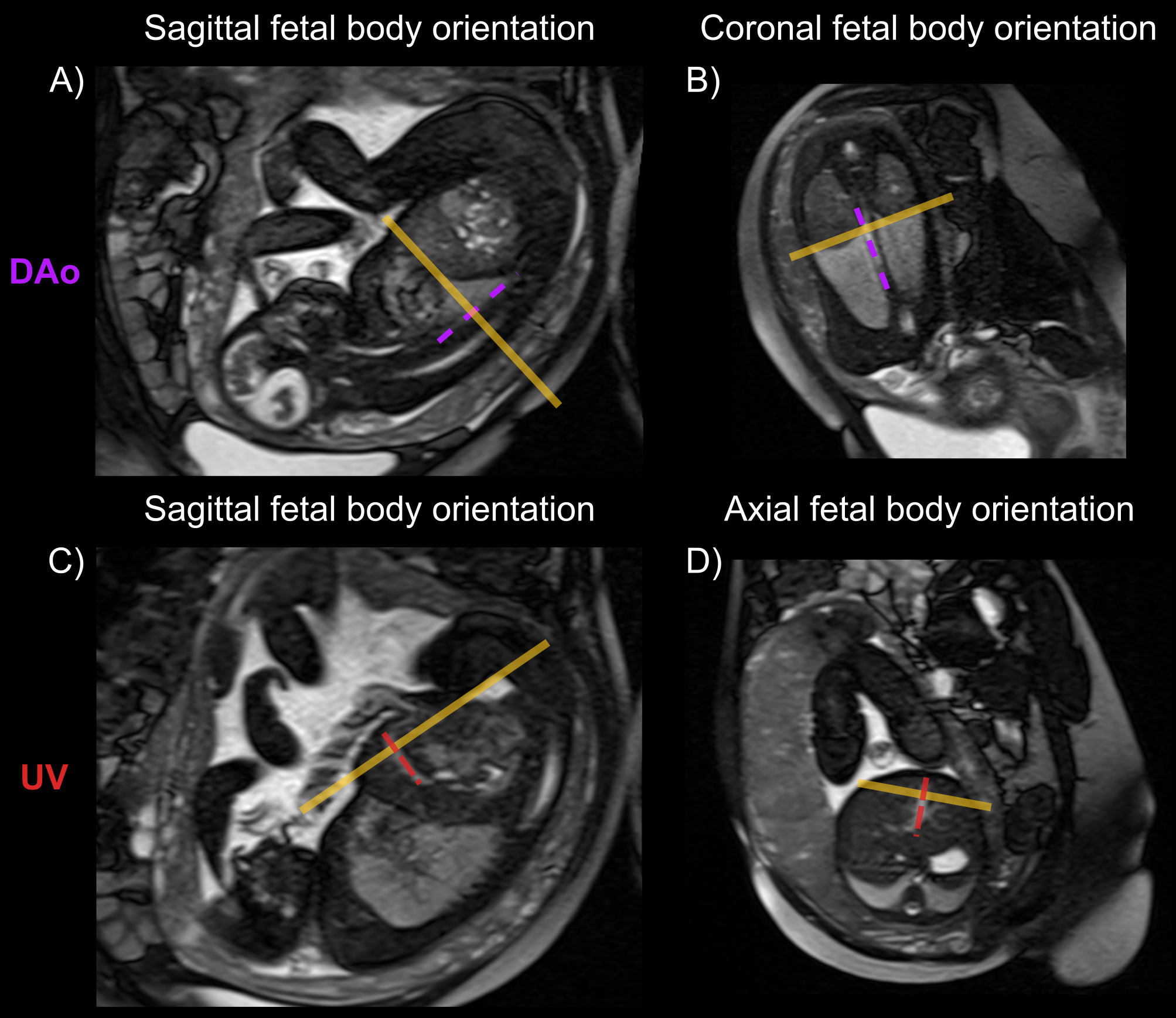}
    \caption{Conventional manual planning of the 2D phase-contrast flow sequences of the descending aorta and umbilical vein using balanced steady-state free precession (bSSFP) acquisitions for visualizing the anatomy and prescribing the planes. A-B illustrate sagittal and coronal fetal body acquisitions with the DAo highlighted (purple dotted line) and the 2D PC acquisition slices (yellow) planned perpendicularly to the vessel. Similarly, C and D illustrate the sagittal and axial bSSFP acquisitions of the fetal body, with the UV highlighted (dotted red line) and the 2D PC planning (yellow lines). Usually, two planes of the bSSFP scans are acquired to ensure the location and orientation of the PC slices are correct.}
    \label{radiographer_planning}
\end{figure*}

\section*{Methods}
\begin{figure*}[t!]
    \centering
    \includegraphics[width=0.75\textwidth]{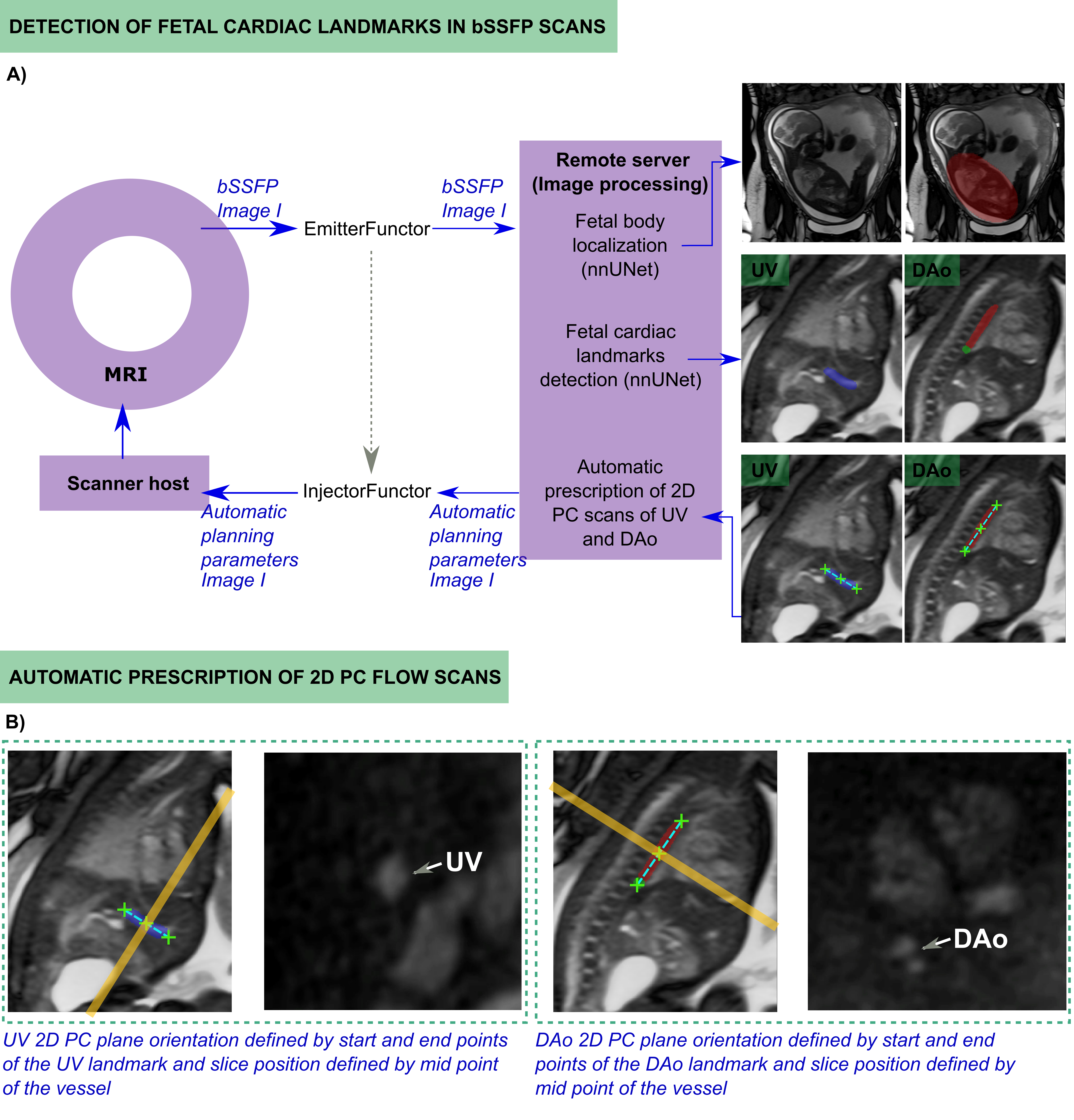}
    \caption{Schematic overview of the entire pipeline for automatic plane prescription. (A) Detection of the fetal cardiac landmarks in the bSSFP sequence and (B) Planning parameters applied in the 2D PC sequences.}
    \label{fig_overview}
\end{figure*}

\noindent The processing workflow for OWL consists of the following steps: (A) The fetal body is automatically extracted on a whole uterus bSSFP scan and a bounding box is calculated, zooming into the thorax; (B) Cardiac landmarks are automatically identified and specific points - e.g., start, mid, and endpoints of the UV segmentation - computed; (C) The orientation and position of the planes cross-sectional to the vessels (that pass through the vessel midpoint) are calculated and applied to the following 2D PC sequences of the UV and DAo. Figure \ref{fig_overview} illustrates the workflow. This method was implemented on a 0.55T scanner (MAGNETOM Free.Max, Siemens Healthcare, Erlangen, Germany).



\subsection*{(A) Localization of the fetal body}
\noindent The OWL workflow relies on a single acquisition of a bSSFP scan covering the whole-uterus in the coronal orientation, acquired in under one minute (see Table \ref{table_parameters} for parameters). Once acquired and reconstructed in the scanner, the bSSFP image was immediately exported to the GPU-powered remote server for processing. This was enabled by the FIRE \cite{chow2021prototyping} framework, installed in the scanner, enabling prospective sending and retrieval of data between the scanner reconstruction environment and an external GPU-equipped (NVIDIA GEFORCE RTX 2080 Ti, NVIDIA Corporate, Santa Clara, CA) computer.

\noindent The images were converted to ISMRMRD format immediately upon acquisition and entered a processing pipeline. This pipeline involves, as an initial step, deep learning-based extraction of the fetal body using a pre-trained 3D nnUNet \cite{Isensee2021}. This framework performs semantic segmentation and it automatically adapts to a given dataset by analysing the provided training data and configuring a matching UNet-based segmentation pipeline. Next, the calculation of a bounding box for zooming into the fetal thorax was performed to enable automated detection of key cardiac landmarks.

\paragraph{Datasets, training and testing} Fetal MRI scans were acquired on 0.55T and 1.5T clinical scanners (MAGNETOM Free.Max and Sola, Siemens Healthineers, Forchheim, Germany) as part of four ethically approved studies performed between 2022 and 2024 at St Thomas' Hospital (miBirth 23/LO/0685, MEERKAT REC19/LO/0852, NANO 19LO0736, and Quantification of Fetal Growth and Development Using MRI 07/H0707/105). A total of 167 (gestational ages 17-40 weeks) whole-uterus coronal multi-echo gradient-echo EPI acquired at 0.55T (114 cases) and a mixture of whole-uterus and sagittal fetal body bSSFP (53 in total, 20 at 1.5T and 33 at 0.55T) scans, acquired as part of research protocols, were used for training the 3D nnUNet fetal body segmentation model. Data augmentation using the TorchIO framework \cite{PerezGarcia2021} (random anisotropy, ghosting, bias field, noise, blur and intensity re-scaling) was applied, resulting in 1837 images. For testing, 43 further datasets (28 0.55T EPI, 10 0.55T bSSFP and 5 1.5T bSSFP) were used. The acquisition parameters can be found in Table \ref{table_parameters}.

\subsection*{(B) Landmark detection}
\noindent A further instance of the nnUNet \cite{Isensee2021} framework was adopted and trained to extract fetal cardiac landmarks - specifically the UV, DAo, the endpoint of the DAo at the diaphragm level (DAo-D), heart-liver interface (H-L), and apices of the lungs (AP-L). Although the current pipeline only processes the UV and DAo automated segmentations for automatic planning, other landmarks have already been integrated into the network training, for instance, for automatizing the planning of the superior vena cava using DAo-D and AP-L, or H-L for modifying the slice positioning of the DAo scan. The start, mid, and endpoints of the DAo and UV segmentations were computed, transformed into the patient coordinate system and written into a text file on the external server and scanner host. Figure \ref{fig_overview}B illustrates the key points extracted and used for planning the PC acquisitions.

\noindent Similarly, to mitigate challenges associated with detecting cardiac vasculature landmarks, particularly in pathological cases or when fetal positioning obscures vessel visibility from a coronal whole-uterus orientation, a further landmark was included - the length of the thoracic spine at the level of the lungs, stabilizing the DAo planning due to its prominent contrast and alignment parallel to the DAo. 

\paragraph{Datasets, training and testing}
For this step, the training of the cardiac landmark detection included 71 cases at 1.5T and 23 at 0.55T. These bSSFP images were mostly oriented to the sagittal fetal body, although some maternal coronal-oriented images, with variable fetal lie, were added to increase robustness. These images were cropped to the fetal thorax, and data augmentation was again applied using the TorchIO framework \cite{PerezGarcia2021}, resulting in 1136 images. The gold standard segmentations of the vessels and planning landmarks were manually drawn for 63 cases (1008 original and augmented images) by a clinician with 3 years of experience in paediatric and fetal cardiology, and the additional spine landmark was segmented on all datasets by a fetal expert with 4 years of experience. The trained model was tested on 4 1.5T and 6 0.55T bSSFP datasets. For all datasets used for training and testing, the inference of the fetal body segmentation network was applied first to the images and the resulting bounding box containing the fetal thorax position was taken for the subsequent steps. 

\begin{table*}[h!]
\centering
\resizebox{\textwidth}{!}{
\begin{tabular}{|l|l|c|c|} 
\rowcolor{gray!40}\textbf{Dataset} & \textbf{Parameters} & \textbf{Subjects} & \textbf{Time} \\
\hline
\rowcolor{white}\textbf{EPI training} 0.55T & 0.55T Siemens MAGNETOM Free.Max & & \\
\rowcolor{white} & 6-channel coil, 9-channel spine coil & & \\
\rowcolor{white} & Matrix=128x128 mm, Resolution 3.1 mm\textsuperscript{3} & & \\
\rowcolor{white} & TE=[46, 120, 194] ms, TR=10.4–18.4 s & & \\
\rowcolor{white} & 50-55 slices, 15-30 repetitions & 114 & 4-6 min \\ 
\hline
\rowcolor{white}\textbf{bSSFP training} 1.5T & 1.5T Siemens MAGNETOM Sola & & \\
\rowcolor{white} & 18- and 30-channel coils & & \\
\rowcolor{white} & Matrix=512x512 mm, Resolution 0.68x0.68x3 mm & & \\
\rowcolor{white} & 35 slices, TE=2.4 ms, TR=439.3 ms & 50 & 30 s \\ 
\hline
\rowcolor{white}\textbf{bSSFP training} 0.55T & 0.55T Siemens MAGNETOM Free.Max & & \\
\rowcolor{white} & 6-channel coil, 9-channel spine coil & & \\
\rowcolor{white} & Matrix=[288-480]x[288-576] mm, Resolution 0.73x0.73x[3-5] mm & & \\
\rowcolor{white} & 25-70 slices, TE=4.0 ms, TR=691.87 ms & 28 & 63 s \\ 
\hline
\rowcolor{gray!20}\textbf{bSSFP autoplan} & 0.55T Siemens MAGNETOM Free.Max & & \\
\rowcolor{gray!20} & 6-channel coil, 9-channel spine coil & & \\
\rowcolor{gray!20} & Matrix=288x288, Resolution 1.46x1.46x4 mm & & \\
\rowcolor{gray!20} & 55 slices, TE=4.22 ms, TR=1108.58 ms & 7 & 56 s \\ 
\hline
\rowcolor{gray!20}\textbf{2D PC autoplan} & 0.55T Siemens MAGNETOM Free.Max & & \\
\rowcolor{gray!20} & 6-channel coil, 9-channel spine coil & & \\
\rowcolor{gray!20} & Matrix=208x208, Resolution 1.44x1.44x5 mm & & \\
\rowcolor{gray!20} & 25 temporal positions, TE=5.31 ms, TR=78.64 ms & 7 & 23-30 s \\ 
\end{tabular}
}
\caption{Sequence parameters for all described sequences used for training (first three rows, white) and for the prospective OWL cases (last two rows, grey).}
\label{table_parameters}
\end{table*}

\subsection*{(C) Calculation of the 2D PC planes}
\subsubsection*{Automatic orientation and slice positioning calculation}
\noindent The 2D PC sequences were modified to access the landmark information stored in a file shared between the remote server and scanner host, which includes the start, mid, and endpoints of the UV and DAo (first two prospective cases) or UV and spine (other 5 prospective cases), to calculate the orientation and slice positioning of 2D planes cross-sectional to these two major vessels. The sequence modifications involved calculating the vector defined by the start and endpoints of each vessel segmentation, which represents the normal vector to the 2D slice orientation (see Figure \ref{fig_overview}B). The midpoint of each vessel was used to define the centre position of the slice, where the indexed flow measurements would be measured.

\subsection*{Experiments and evaluation}

\subsubsection*{Fetal body extraction and cardiac landmark detection models}
\noindent The performance of the fetal body segmentation network was analyzed using the Dice similarity coefficient (DSC) and the Intersection-over-Union (IoU) metric calculated between the gold standard manual segmentations performed by two fetal experts (6 and 4 years of fetal MRI experience, respectively) and the body predictions generated by the network to allow comparison. The cardiac landmark detection performance was analyzed by calculating the 3D distance between the centre-of-mass (CoM) of the manual segmentations performed by a paediatric cardiologist and a fetal expert (3 and 4 years of experience, respectively) and the segmentations obtained from the network.

\subsubsection*{Real-time fetal vessel planning}
\noindent The entire pipeline was acquired prospectively in 7 pregnant volunteers in St Thomas’ Hospital, recruited in June-July 2024 after informed consent was obtained as part of three ethically approved studies (miBirth 23/LO/0685, MEERKAT REC19/LO/0852, NANO 19/LO/0736). Women were scanned on the above-described clinical 0.55T MAGNETOM Free.Max scanner in the supine position with leg support to ensure comfort. Frequent verbal interaction and life monitoring were performed throughout the scan. Gestational ages ranged between 36+3 and 39+6 weeks. For each fetal subject, the initial whole-uterus coronal anatomical bSSFP scan was acquired. Subsequently, the automatically-planned 2D PC scans (2D PC autoflow) research sequence was acquired for the UV and DAo vessels with optimized parameters for 0.55T \cite{Cleri2024}. All sequence parameters can be found in Table \ref{table_parameters}. In addition, the same flow scans were manually planned and acquired by fetal radiographers (1-5 years of experience) for all subjects. For each fetal subject, 1-3 automatic scans were acquired for each vessel, and 1-2 manual acquisitions. The time required for planning by the radiographers and the automatic method was measured.

\noindent The planning quality was assessed with a scoring system that evaluated the position and orientation of the resultant acquisition plane. The position refers to the point along the vessel at which the plane intersects the vessel of interest, and the orientation refers to whether the plane was adequately perpendicular to the direction of blood flow. The scoring system was as follows: 2 - no adjustments were considered necessary, 1 - minor adjustments were desirable but the planning was still appropriate for quantification of blood flow, and 0 - the resultant sequence was deemed inadequate for quantitative assessment due to planning. The maximum score for each scan is therefore 4 for optimal position and orientation planning.

\subsubsection*{Fetal blood flow measurements from automated acquisitions}
\noindent To evaluate the feasibility of using automatically planned PC sequences for fetal blood flow measurements, 19 automatic PC sequences targeting the UV (10) and DAo (9) vessels were acquired from 7 fetal subjects, and 17 (11 UV and 6 DAo) manual scans were planned and acquired for comparison. All acquisitions were visually assessed for motion artefacts and incorrect planning. The datasets deemed of sufficient quality for flow analysis were subsequently retrospectively gated using metric optimized gating (MOG) \cite{jansz2010metric}. Quantitative flow measurements were then conducted using the cardiac tool cvi42 V5.11 (Circle Cardiovascular Imaging Inc. Calgary, Canada), with the flow curves subsequently plotted. Fetal flow measurements were indexed to estimated fetal weight to enable comparison of the flow (in ml/min) to published reference ranges \cite{prsa2014}.

\noindent Fetal weight was estimated by obtaining the fetal volume by segmenting the fetal body from a bSSFP image of the maternal uterus using \cite{Uus2021UNET} and converting volume to weight \cite{barker1994,prsa2014} as estimated fetal weight: 
\begin{equation}
EFW = \textrm{fetal volume (mL)} * 1.031 + 120
\end{equation}

\section*{Results}
\noindent Both retrospective and prospective results for the proposed OWL method are presented in the following sections, with the fetal body and the landmark detection analysis performed retrospectively and the automatic planning assessment performed prospectively.

\subsection*{(A) Extracting the fetal body}
\begin{figure*}[ht]
    \centering
    \includegraphics[width=.85\textwidth]{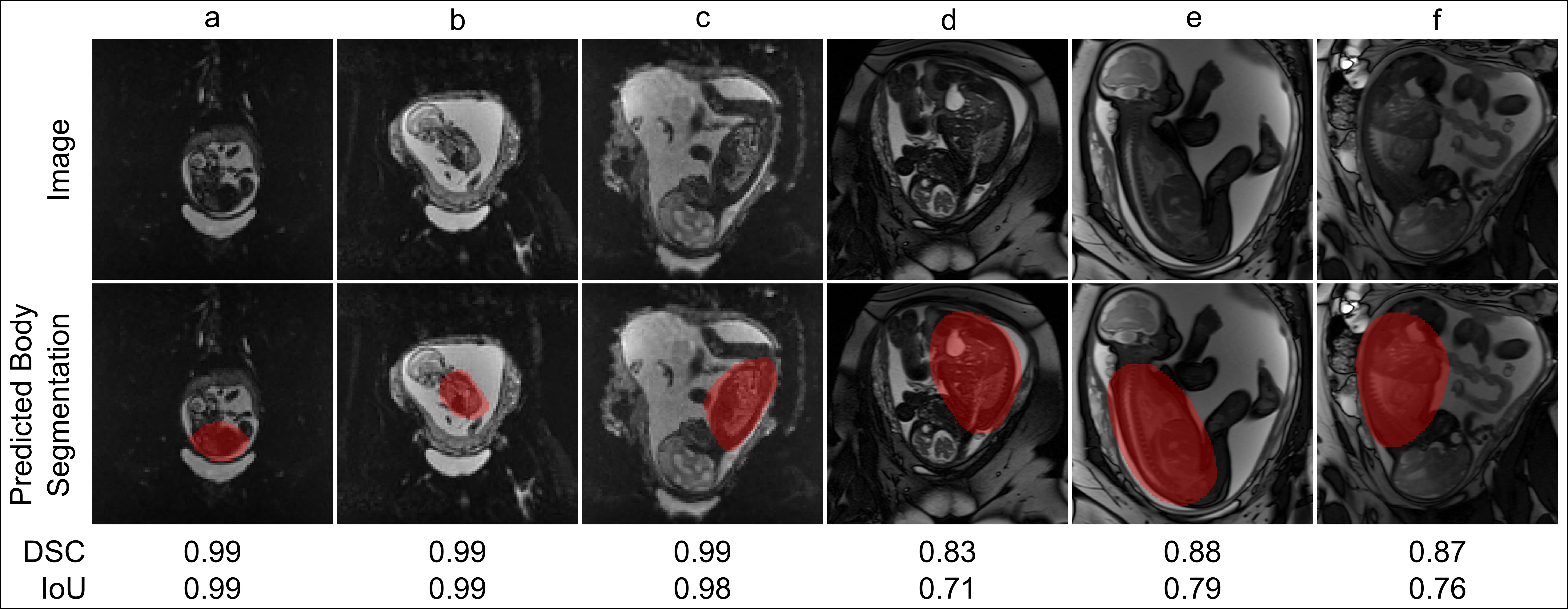}
    \caption{Fetal body automated segmentations for six fetal subjects of the test set: (a-c) extracted from EPI and (d-f) from bSSFP scans.}
    \label{fig_body_results}
\end{figure*}

\noindent The fetal body segmentation task, trained on 0.55T and 1.5T data, achieved an overall DSC of 0.94$\pm$0.05 and IoU of 0.9$\pm$0.08. Extraction of the fetal body took 6.68 seconds per volume in the offline testing mode. Figure \ref{fig_body_results} illustrates the fetal body localization results for 6 fetal subjects and the respective DSC and IoU computed.

\subsection*{(B) Fetal cardiac landmark detection}
\begin{figure*}[ht]
    \centering
    \includegraphics[width=.85\textwidth]{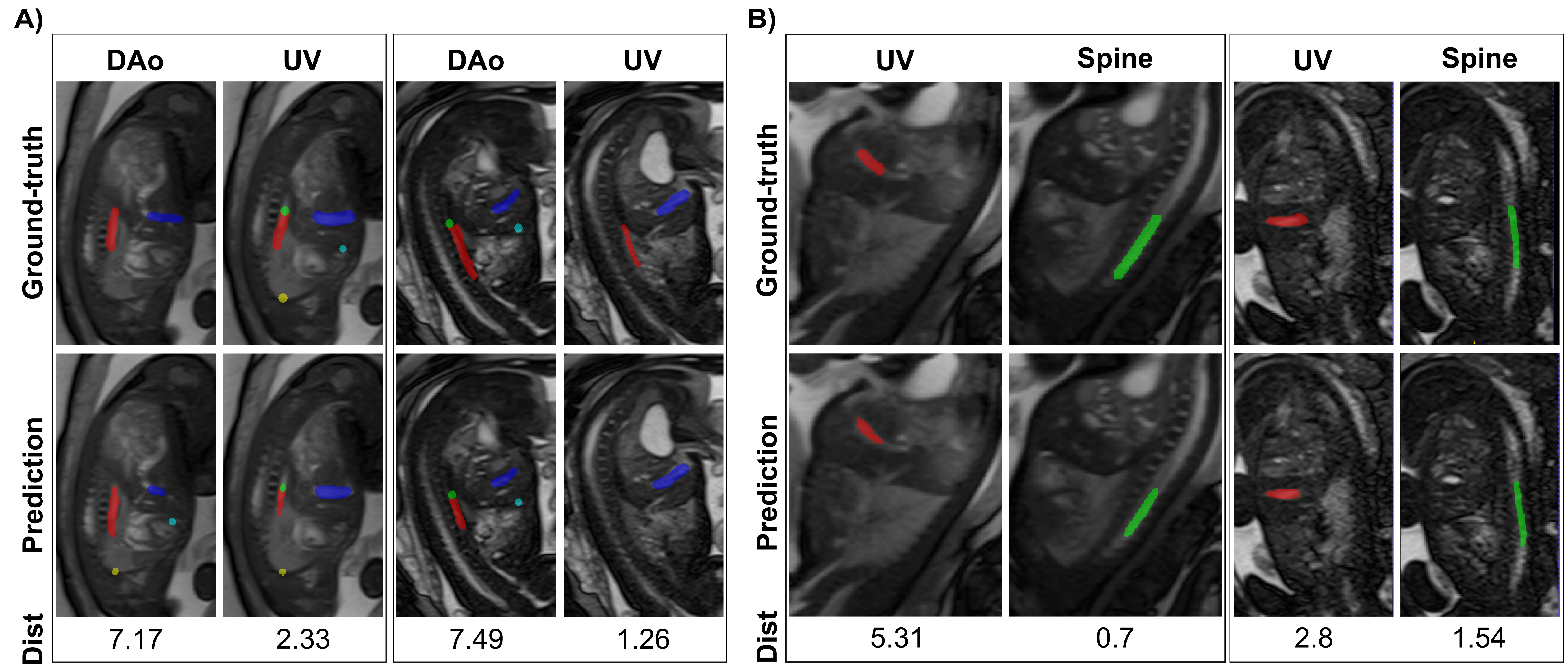}
    \caption{Ground-truth and predicted cardiac landmarks, as well as distances between center-of-mass coordinates of manual and automated segmentations, are depicted for the three landmarks of interest - DAo, UV, and spine. In (A), the UV (blue) and DAo (red) segmentations are shown with the respective distances calculated for these two labels. Other landmarks are also depicted and will be used in the extension and optimization of OWL. (B) shows the results of the spine (green) and UV (red) detection.}
    \label{fig_landmarks}
\end{figure*}

\noindent The mean distance between the CoM of the landmarks in the automatic and manual segmentations was 5.77$\pm$2.91mm, 4.32$\pm$2.44 mm and 4.94$\pm$3.82 mm for the DAo, spine and UV, respectively. Examples can be found in Figure \ref{fig_landmarks}, with the predicted landmarks generated by the nnUNet, the corresponding ground-truth segmentations, and the distance between the two depicted for 4 fetal subjects of the test set. Landmark detection took 5.86 seconds per volume in the offline testing mode.

\subsection*{(C) Real-time planning of PC scans of the major vessels}
\begin{figure*}[ht]
    \centering
    \includegraphics[width=.8\textwidth]{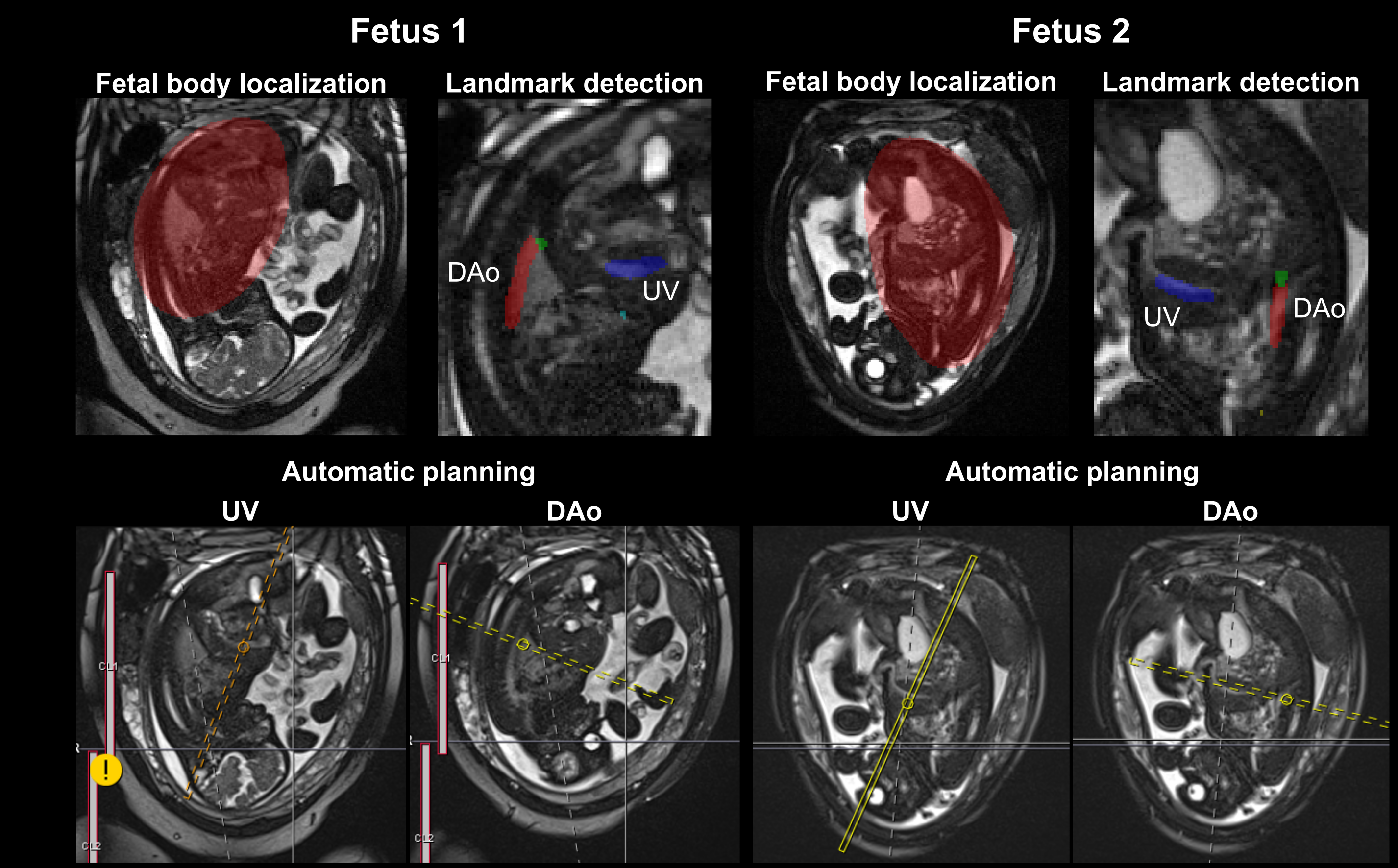}
    \caption{The two-step localization approach of fetal body and cardiac landmarks is depicted for two fetal subjects, as well as the automatic 2D PC slice orientation and positioning based on the detected descending aorta and umbilical vein segmentations.}
    \label{fig_prospective}
\end{figure*}

\begin{figure*}[t!]
    \centering
    \includegraphics[width=.85\textwidth]{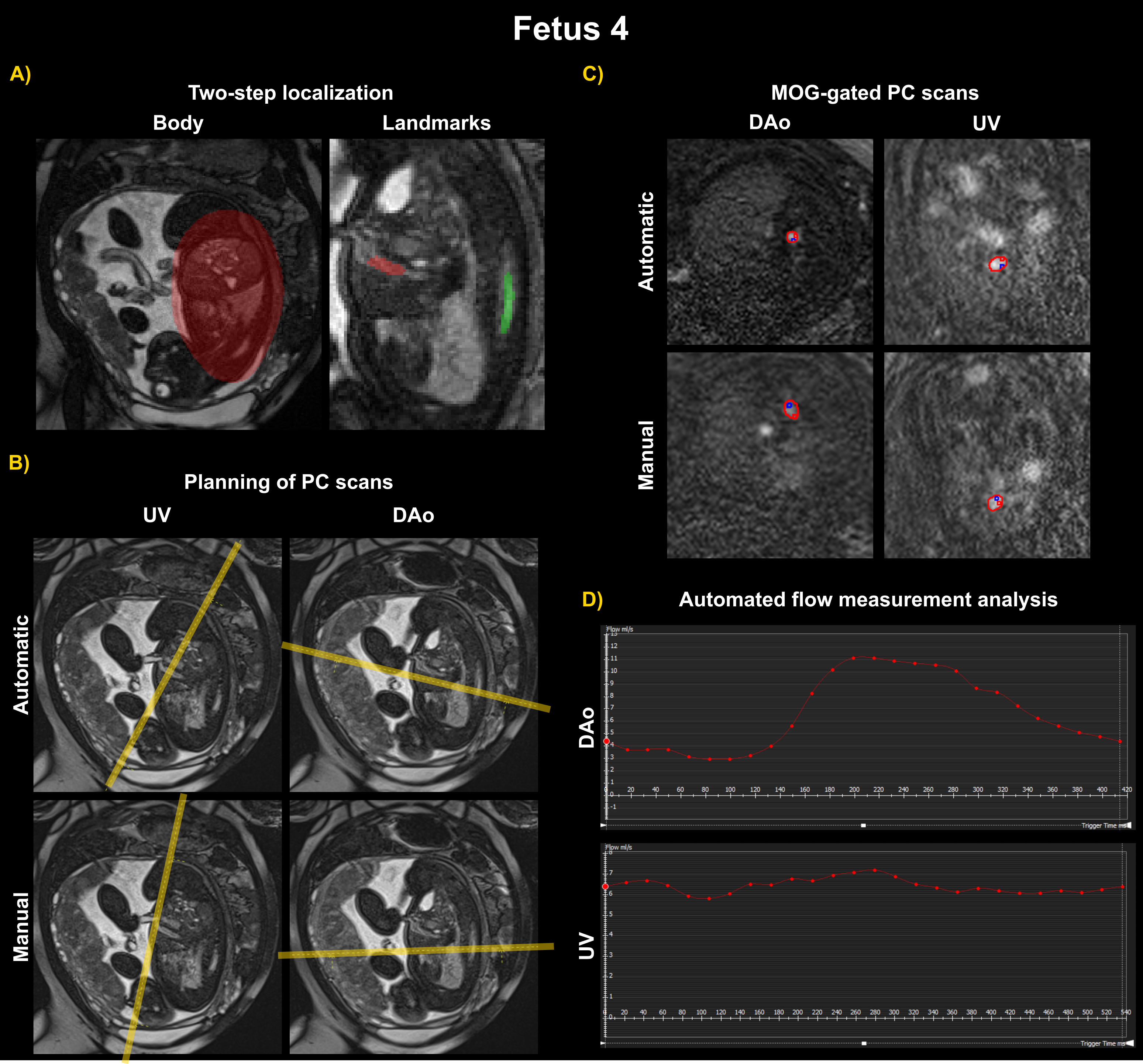}
    \caption{All steps of the automatic planning pipeline are shown for fetal subject 4 of the prospective dataset: 1) two-step localization of the fetal body, followed by the cardiac landmarks where points of interest are extracted from for planning (A); 2) automatic planning of the umbilical vein and descending aorta using the respective extracted landmarks (B); 3) MOG motion-corrected 2D PC scans acquired using the automatic workflow and manual planning, for comparison (C); 4) analysis of the flow measurements extracted from the automated and manual acquisitions, for validation of the developed framework (D).}
    \label{fig_analysis}
\end{figure*}

\noindent The entire pipeline was successfully prospectively run in 6 out of 7 fetal scans. For the first two fetal subjects, the DAo and UV landmarks were used for guiding planning. Figure \ref{fig_prospective} depicts the real-time cardiac landmark detection results obtained from the two-step localization task, followed by the automatic planning of the 2D PC scans in the scanner console. In the following eight subjects, the spine and UV landmarks were used. Figure \ref{fig_analysis} depicts the entire processing pipeline for one exemplary prospective case. A) shows the results of the two-step localization pipeline, B) the automatic planning, based on the real-time detected cardiac landmarks, and matching manual planning, C) the MOG-gated automatic and manual 2D PC acquisitions, and D) the resulting flow measurement analysis from the gated automated scans. The estimated time between the completion of the bSSFP sequence and the start of the 2D PC sequence, ready to be run with the planning parameters correctly set for the two sequences, was 18.25 seconds. The average manual planning time for both sequences by a paediatric cardiologist varied between 1-2 minutes, and for a fetal specialist radiographer 2:30 minutes on average, although late gestation and challenging fetal positions may require an additional localization sequence that increases planning time up to 5 minutes. 

\subsection*{(C) Qualitative and quantitative analysis of the PC scans}
\noindent The overall planning quality scores for the automatic versus manual 2D PC acquisitions, performed by a fetal specialist radiographer (automatic vs. manual), were as follows: DAo - 3.5/4 vs. 3.25/4; UV - 2.43/4 vs. 2.86/4. The overall planning quality was 2.73/4 vs. 3/4. Position and orientation were also assessed individually (automatic vs. manual), with results as follows: position - 1.36/2 vs. 1.45/2; orientation - 1.36/2 vs. 1.55/2.

\noindent From the seven prospective cases, the landmark detection task failed for subject 5 due to a challenging fetal lie unfamiliar to the network, and thus it was disregarded from any further analysis. For subjects 1 and 2, the PC sequence parameters used were not yet optimised, thus image quality was poor despite good planning (3.0/4 for each case), and MOG gating and analysis were not performed. A total of 16 PC sequences from four fetuses were successfully gated with MOG. Ten were acquired using automatic planning and six were manually planned. Three automatically planned sequences were not suitable for quantitative flow assessment due to the oblique intersection of the vessel of interest (two for the DAo and one for the UV). The remaining 13 PC sequences are detailed in Table 2. Planning quality assessment was additionally performed by a clinician (paediatric cardiologist with fetal experience) on the 13 MOG-gated acquisitions where flow analysis was performed (7 automatic and 6 manual) and, overall, the planning scores (automatic vs. manual) were 1.92/2 vs. 2/2. 

\begin{table*}[!ht]
    \centering
    \begin{tabular}{|c|c|c|l|c|c|c|c|}
    \hline
        \rowcolor{gray!40}\textbf{Case} & \textbf{EFW} & \textbf{Vessel} & \textbf{Method} & \textbf{Flow} & \textbf{Indexed flow} & \textbf{Reference range\cite{prsa2014}} & \textbf{$\Delta$} \\ \hline
        \rowcolor{gray!40}~ & (kg) & ~ & ~ & (ml/min) & (ml/kg/min) & (ml/kg/min) & (\%) \\ \hline
        \rowcolor{white}\textbf{3} & 3.47 & DAo & Automated & 670 & 193 & (160 - 344) & ~ \\ \hline
        \rowcolor{gray!20}~ & ~ & ~ & Manual & 470 & 197 & ~ & ~ \\ \cline{4-6} \cline{8-8}
        \rowcolor{gray!20}~ & ~ & ~ & Automated & 400 & 168 & ~ & 14.9 \\ \cline{4-6} \cline{8-8}
        \rowcolor{gray!20}~ & ~ & \multirow[l]{-3}{*}{DAo} & Automated & 457 & 192 & \multirow[c]{-3}{*}{(160 - 344)} & 2.8 \\ \cline{3-8}
        \rowcolor{gray!20}~ & ~ & ~ & Manual & 337 & 142 & ~ & ~ \\ \cline{4-6} \cline{8-8}
        \rowcolor{gray!20}\multirow[c]{-5}{*}{\textbf{4}} & \multirow[c]{-5}{*}{2.38} & \multirow[l]{-2}{*}{UV} & Automated & 385 & 162 & \multirow[c]{-2}{*}{(62 - 206)} & -14.2 \\ \hline
        \rowcolor{white}~ & ~ & DAo & Manual & 729 & 290 & (160 - 344) & ~ \\ \cline{3-8}
        \rowcolor{white}~ & ~ & ~ & Manual & 375 & 149 & ~ & ~ \\ \cline{4-6} \cline{8-8}
        \rowcolor{white}\multirow[c]{-3}{*}{\textbf{6}} & \multirow[c]{-3}{*}{2.52} & \multirow[c]{-2}{*}{UV} & Automated & 411 & 163 & \multirow[c]{-2}{*}{(62 - 206)} & -9.6 \\ \hline
        \rowcolor{gray!20}~ & ~ & ~ & Manual & 928 & 361 & ~ & ~ \\ \cline{4-6} \cline{8-8}
        \rowcolor{gray!20}~ & ~ & \multirow[c]{-2}{*}{DAo} & Automated & 984 & 383 & \multirow[c]{-2}{*}{(160 - 344)} & -6.0 \\ \cline{3-8}
        \rowcolor{gray!20}~ & ~ & ~ & Manual & 402 & 156 & ~ & ~ \\ \cline{4-6} \cline{8-8}
        \rowcolor{gray!20}\multirow[c]{-4}{*}{\textbf{7}}  & \multirow[c]{-4}{*}{2.57} & \multirow[c]{-2}{*}{UV} & Automated & 397 & 154 & \multirow[c]{-2}{*}{(62 - 206)} & 1.2 \\ \hline
    \end{tabular}
    \caption{Quantitative flow data for each fetus where automated planning was applied. EFW = estimated fetal weight, $\Delta$ = the percentage difference between automatically planned flow measurements and their paired manually planned measurements. Fetal subjects 3, 4, 6, and 7 of the prospective dataset were included in this analysis.}
    \label{table_flow_data}
\end{table*}


\section*{Discussion and Conclusion}
\noindent This study presents the fully automated OWL method for the planning of 2D PC sequences of two major vessels in the fetal circulation - the umbilical vein and the descending aorta. The pipeline uses a single anatomical whole-uterus coronal bSSFP scan to extract specific vascular, cardiac and fetal body landmarks that are immediately transferred to the subsequent PC sequences to plan the acquisition planes orthogonal to the vessels, with the entire processing pipeline running in less than 20 seconds. The scans obtained automatically showed comparable performance to the manually planned acquisitions - with no significant differences in indexed fetal flow measurements - thus carrying the potential to reduce operator dependence and increase time efficiency in the acquisition of such complex scans, thereby widening access to crucial information for cardiac diagnostics beyond specialist centres. The segmented landmarks were successfully used in the determination of the 2D PC sequence orientation and slice positioning. Thus, while based on previously shown detection of cardiac landmarks \cite{Xue2021,ramirez2022automated} and automated planning of cardiac scans \cite{Frick2011}, a novel method was successfully developed bespoke to the challenging area of fetal cardiac MRI. All processing tools were successfully implemented and prospectively tested in real-time.

\noindent The retrospective evaluation of fetal body localization and landmark extraction showed strong agreement with ground-truth segmentations performed by fetal and paediatric experts, evidenced by high DSC and IoU scores in the body extraction task, and minimal distances between the CoM of the manually segmented and automatically extracted landmarks. These results indicate sufficient robustness for the accurate calculation of the 2D PC planes. Prospective results showed notable agreement between automatic planning and manual planning in terms of planning quality and derived quantitative flow results, with significantly reduced planning time.

\noindent When scoring the adequacy of the PC sequence planning, 7 automatic and 6 manual sequences were deemed suitable for flow analysis. The majority of indexed fetal flow measurements were within the reference range. The indexed DAo flow was slightly higher than the reference range for fetus 7, however this finding was present in both the automated and manually planned PC sequences. When indexed fetal flow measurements were obtained from both planning methods within the same vessel there was a mean difference in indexed flow of -1.8\% (range -14.2\% to 14.9\%) with no obvious bias present in this small, initial sample. The differences observed in these cases are consistent with measurements taken within the same vessel in other published data \cite{ryd2019}, suggesting the difference in planning methodology did not lead to important differences in quantitative flow assessment.

\noindent Major strengths of the current study and the developed OWL method are the complete and robust real-time deployment, shown prospectively on 7 fetuses, the speed (below 20 seconds), as well as the acquisition of a single bSSFP anatomical scan to visualize, extract and calculate the planes, instead of the at least three bSSFP scans in different orientations of the fetal body employed in current practice.

\noindent All steps of the OWL pipeline - extraction of the fetal body and cardiac landmarks, and calculation of the points of interest in the scanner coordinate system - are available to all interested researchers (https://github.com/saranevessilva/fetal-cardiac-landmarks) and have successfully been integrated into the FIRE setup \cite{chow2021prototyping}, a key step to widen accessibility. Anonymized fetal data, both the bSSFP, the PC flow datasets and the segmentations, are available upon request.

\noindent While the pipeline was applied here on low-field (0.55T) fetal MRI data, the OWL method is not limited to this field strength. Given the excellent results obtained even in the low-field data with low SNR, the use of 1.5T fetal MRI datasets during the training are encouraging for generalization to all field strengths.

\noindent There are, however, limitations with the current study. First, the OWL method was applied to a small number of fetal subjects prospectively, thus the framework will next be applied to a larger and wider cohort and across different scanners/field strengths and pathologies. Additionally, the use of vessel-based landmarks might require significant adaptation, particularly for vessels in more challenging orientations and in CHD pathologies where the major vessels may be of different sizes, positions and orientations. The integration of landmarks as independent as possible from the cardiac anatomy could complement the current vessel-based landmarks as a beneficial next step. Furthermore, the study was limited to the third trimester. Earlier GAs and the associated significant amounts of motion hinder the continuity between the bSSFP and flow scans - a limitation however even aggravated by the additional delay involved in manual planning.

\noindent This study demonstrates the feasibility of rapid, deep learning-based automatic planning for 2D PC scans of major fetal circulation vessels and introduces a framework open to further extensions and improvements. Future directions involve developing a fully self-guided fetal MRI scan, where all aspects of image acquisition and analysis are automated. This aims to facilitate easy dissemination and integration of advanced and complex techniques in non-specialist centres, ultimately widening access to fetal MRI worldwide. Future work includes the addition of automated quality control of the acquisitions to trigger re-calculation and re-acquisition of the sequences, as well as automated analysis with results available by the end of the scan.

\section*{Acknowledgements}
\noindent The authors thank all pregnant women and their families for taking part in this study, the midwife Chidinma Iheanetu Oguejiofor and clinical research fellows Simi Bansal, Hadi Waheed and Vanessa Kyriakopoulou for their invaluable efforts in recruiting and looking after the women in this study as well as Chester John Fauni and Wendy Norman for their involvement in the acquisition of these datasets. This work was supported by a Wellcome Trust Collaboration in Science grant [WT201526/Z/16/Z], Heisenberg funding from the DFG [502024488], a UKRI FL fellowship, an NIHR Advanced Fellowship [NIHR3016640], an EPSRC Research Council DTP grant [EP/R513064/1], the MRC grants [MR/W019469/1] and [MR/X010007/1], and the Wellcome/EPSRC Centre [WT203148/Z/16/Z]. The views presented in this study represent those of the authors and not of Guy's and St Thomas' NHS Foundation Trust.


\section*{Abbreviations}

\noindent \textbf{OWL} AutOmatic floW planning for fetaL MRI  \\
\noindent \textbf{CHD} Congenital heart disease \\
\noindent \textbf{US} Ultrasound \\
\noindent \textbf{MR} Magnetic resonance \\
\noindent \textbf{MRI} Magnetic resonance imaging \\
\noindent \textbf{PC} Phase-contrast \\
\noindent \textbf{BMI} Body mass index \\
\noindent \textbf{TE} Echo time \\
\noindent \textbf{SNR} Signal-to-noise ratio \\
\noindent \textbf{CNN} Convolutional neural network \\
\noindent \textbf{DAo} Descending aorta \\
\noindent \textbf{UV} Umbilical vein \\
\noindent \textbf{bSSFP} Balanced steady-state free precession \\
\noindent \textbf{FIRE} Framework for image reconstruction environment \\
\noindent \textbf{EPI} Echo-planar imaging \\
\noindent \textbf{DSC} Dice similarity coefficient \\ 
\noindent \textbf{IoU} Intersection-over-union \\
\noindent \textbf{CoM} Centre-of-mass \\
\noindent \textbf{MOG} Metric optimized gating \\ 
\noindent \textbf{EFW} Estimated fetal weight \\


\newpage
\begin{thebibliography}{30}
\providecommand{\natexlab}[1]{#1}
\providecommand{\url}[1]{\texttt{#1}}
\providecommand{\urlprefix}{}

\bibitem[{Ferrazzi et~al.(2000)Ferrazzi, E. and Rigano, S. and Bozzo, M. and Bellotti, M. and Giovannini, N. and Galan, H. and Battaglia, F. C.}]{Ferrazzi2000}
Ferrazzi E, Rigano S, Bozzo M, Bellotti M, Giovannini N, Galan H, et~al.
\newblock Umbilical vein blood flow in growth-restricted fetuses.
\newblock Ultrasound in Obstetrics \& Gynecology 2000;16(5):432--438.
\newblock \urlprefix\url{https://obgyn.onlinelibrary.wiley.com/doi/abs/10.1046/j.1469-0705.2000.00208.x}.

\bibitem[{Browne et~al.(2015)Browne, Vaughn A. and Julian, Colleen G. and Toledo-Jaldin, Lillian and Cioffi-Ragan, Darleen and Vargas, Enrique and Moore, Lorna G.}]{Browne2015}
Browne VA, Julian CG, Toledo-Jaldin L, Cioffi-Ragan D, Vargas E, Moore LG.
\newblock Uterine artery blood flow, fetal hypoxia and fetal growth.
\newblock Philosophical Transactions of the Royal Society B: Biological Sciences 2015;370(1663):20140068.
\newblock \urlprefix\url{https://royalsocietypublishing.org/doi/abs/10.1098/rstb.2014.0068}.

\bibitem[{Rurak et~al.(2011)Rurak, Dan and Lim, Ken and Sanders, Ari and Brain, Ursula and Riggs, Wayne and Oberlander, Tim F}]{Rurak2011}
Rurak D, Lim K, Sanders A, Brain U, Riggs W, Oberlander TF.
\newblock Third trimester fetal heart rate and Doppler middle cerebral artery blood flow velocity characteristics during prenatal selective serotonin reuptake inhibitor exposure.
\newblock Pediatr Res 2011 Jul;70(1):96--101.

\bibitem[{Qasim et~al.(2024)Qasim, Amna and Morris, Shaine A and Belfort, Michael A and Qureshi, Athar M}]{Qasim2024}
Qasim A, Morris SA, Belfort MA, Qureshi AM.
\newblock Current understanding of indications, technical aspects and outcomes of fetal cardiac interventions.
\newblock Interv Cardiol Clin 2024 Jul;13(3):319--331.

\bibitem[{McQuillen et~al.(2010)McQuillen, Patrick S and Goff, Donna A and Licht, Daniel J}]{McQuillen2010}
McQuillen PS, Goff DA, Licht DJ.
\newblock Effects of congenital heart disease on brain development.
\newblock Prog Pediatr Cardiol 2010 Aug;29(2):79--85.

\bibitem[{Nayak et~al.(2015)Krishna S. Nayak and Jon-Fredrik Nielsen and Matt A. Bernstein and Michael Markl and Peter {D. Gatehouse} and Rene {M. Botnar} and David Saloner and Christine Lorenz and Han Wen and Bob {S. Hu} and Frederick H. Epstein and John {N. Oshinski} and Subha V. Raman}]{Nayak2015}
Nayak KS, Nielsen JF, Bernstein MA, Markl M, {D  Gatehouse} P, {M  Botnar} R, et~al.
\newblock Cardiovascular magnetic resonance phase contrast imaging.
\newblock Journal of Cardiovascular Magnetic Resonance 2015;17(1):71.
\newblock \urlprefix\url{https://www.sciencedirect.com/science/article/pii/S1097664723008967}.

\bibitem[{Jiang et~al.(2015)Jiang, Jing and Kokeny, Paul and Ying, Wang and Magnano, Chris and Zivadinov, Robert and Mark Haacke, E}]{Jiang2015}
Jiang J, Kokeny P, Ying W, Magnano C, Zivadinov R, Mark~Haacke E.
\newblock Quantifying errors in flow measurement using phase contrast magnetic resonance imaging: comparison of several boundary detection methods.
\newblock Magn Reson Imaging 2015 Feb;33(2):185--193.

\bibitem[{Morton et~al.(2017)Morton, Paul and Ishibashi, Nobuyuki and Jonas, Richard}]{Morton2017}
Morton P, Ishibashi N, Jonas R
\newblock Neurodevelopmental Abnormalities and Congenital Heart Disease
\newblock Circulation Research 2017 120(6):960--977.

\bibitem[{Gowland et~al.(1998)Gowland, P. A. and Freeman, A. and Issa, B. and Boulby, P. and Duncan, K. R. and Moore, R. J. and Baker, P. N. and Bowtell, R. W. and Johnson, I. R. and Worthington, B. S.}]{gowland_vivo_1998}
Gowland PA, Freeman A, Issa B, Boulby P, Duncan KR, Moore RJ, et~al.
\newblock In vivo relaxation time measurements in the human placenta using echo planar imaging at 0.5 {T}.
\newblock Magnetic Resonance Imaging 1998 Apr;16(3):241--247.

\bibitem[{Chow et~al.(2021)Chow, K and Kellman, P and Xue, H}]{chow2021prototyping}
Chow K, Kellman P, Xue H.
\newblock Prototyping image reconstruction and analysis with FIRE.
\newblock In: SCMR 24th annual scientific sessions. Virtual Meeting; 2021. 

\bibitem{Rodriguez2012}
Rodríguez, M. R., de Vega, V. M., Alonso, R. C., Arranz, J. C., Ten, P. M., Pedregosa, J. P.
\newblock MR imaging of thoracic abnormalities in the fetus.
\newblock Radiographics, 32(7), 2012.

\bibitem{Furey2016}
Furey, E. A., Bailey, A. A., Twickler, D. M.
\newblock Fetal MR imaging of gastrointestinal abnormalities.
\newblock Radiographics, 36(3):904--917, 2016.

\bibitem{Gat2016}
Gat, I., Hoffmann, C., Shashar, D., Yosef, O. B., Konen, E., Achiron, R., Brandt, B., Katorza, E.
\newblock Fetal brain MRI: novel classification and contribution to sonography.
\newblock Ultraschall in der Medizin-European Journal of Ultrasound, 37(2):176--184, 2016.

\bibitem{Xue2019}
Xue, H., Davies, R., Hansen, D., Tseng, E., Fontana, M., Moon, J. C., Kellman, P.
\newblock Gadgetron inline AI: Effective model inference on MR scanner.
\newblock In \textit{Proceedings of the 27th Annual ISMRM Meeting and Exhibition}, page 4837, 2019.

\bibitem{Xue2021}
Xue, H., Artico, J., Fontana, M., Moon, J. C., Davies, R. H., Kellman, P.
\newblock Landmark detection in cardiac MRI by using a convolutional neural network.
\newblock Radiology: Artificial Intelligence, 3(5):e200197, 2021.

\bibitem{Frick2011}
Frick, M., Paetsch, I., den Harder, C., Kouwenhoven, M., Heese, H., Dries, S., Schnackenburg, B., de Kok, W., Gebker, R., Fleck, E., et al.
\newblock Fully automatic geometry planning for cardiac MR imaging and reproducibility of functional cardiac parameters.
\newblock Journal of Magnetic Resonance Imaging, 34(2):457--467, 2011.

\bibitem{ramirez2022automated}
Ramirez Gilliland, P., Uus, A., van Poppel, M. P. M., Grigorescu, I., Steinweg, J. K., Lloyd, D. F. A., Pushparajah, K., King, A. P., Deprez, M.
\newblock Automated multi-class fetal cardiac vessel segmentation in aortic arch anomalies using T2-weighted 3D fetal MRI.
\newblock In \textit{International Workshop on Preterm, Perinatal and Paediatric Image Analysis}, pages 82--93, 2022, Springer.

\bibitem{Cleri2024}
Cleri, M., Woodgate, T., Zhang, C., McElroy, S., Giles, S., Story, L., Pushparajah, K., Lloyd, D., Hutter, J., Payette, K.
\newblock Fetal Blood Flow Measurements at Low Field (0.55T) using Metric Optimized Gating.
\newblock In \textit{Proceedings of the International Society for Magnetic Resonance in Medicine (ISMRM) Annual Meeting}, 2024.

\bibitem{jansz2010metric}
Jansz, M. S., Seed, M., Van Amerom, J. F. P., Wong, D., Grosse-Wortmann, L., Yoo, S.-J., Macgowan, C. K.
\newblock Metric optimized gating for fetal cardiac MRI.
\newblock Magnetic Resonance in Medicine, 64(5):1304--1314, 2010.
\newblock Wiley Online Library.

\bibitem{prsa2014}
Prsa, M., Sun, L., van Amerom, J., Yoo, S., Grosse-Wortmann, L., Jaeggi, E., Macgowan, C., Seed, M.
\newblock Reference ranges of blood flow in the major vessels of the normal human fetal circulation at term by phase-contrast magnetic resonance imaging
\newblock Circulation: Cardiovascular Imaging, 7(4):663-70, 2014.

\bibitem{barker1994}
Baker, P.N., Johnson, I.R., Gowland, P.A., Hykin, J., Harvey, P.R., Freeman, A., Adams, V., Mansfield, P. and Worthington, B.S.  
\newblock Fetal weight estimation by echo-planar magnetic resonance imaging
\newblock The Lancet, 343(8898), pp.644-645, 1994.

\bibitem{ryd2019}
Ryd, D., Sun, L., Steding-Ehrenborg, K., Bidhult, S., Kording, F., Ruprecht, C., Macgowan, C., Seed, M., Aletras, A., Arheden, H., Hedström, E.
\newblock Quantification of blood flow in the fetus with cardiovascular magnetic resonance imaging using Doppler ultrasound gating: validation against metric optimized gating
\newblock Journal of Cardiovascular Magnetic Resonance, 21: 74 ,2019

\newblock \textit{Ultrasound in Obstetrics \& Gynecology}.
\newblock 53(5), 669--675.
\newblock \url{https://doi.org/10.1002/uog.20167}.

\bibitem{Sousa2019}
Tavares de Sousa, M., Hecher, K., Yamamura, J., Kording, F., Ruprecht, C., Fehrs, K., Behzadi, C., Adam, G., Schoennagel, B. P. (2019).
\newblock Dynamic fetal cardiac magnetic resonance imaging in four-chamber view using Doppler ultrasound gating in normal fetal heart and in congenital heart disease: comparison with fetal echocardiography.
\newblock \textit{Ultrasound in Obstetrics \& Gynecology}.
\newblock 53(5), 669--675.
\newblock \url{https://doi.org/10.1002/uog.20167}.

\bibitem{ponrartana2023low}
Ponrartana, S., Nguyen, H. N., Cui, S. X., Tian, Y., Kumar, P., Wood, J. C., Nayak, K. S. (2023).
\newblock Low-field 0.55 T MRI evaluation of the fetus.
\newblock \textit{Pediatric Radiology}.
\newblock 53(7), 1469--1475.
\newblock \url{https://doi.org/10.1007/s00247-023-05619-2}.

\bibitem{aviles2023reliability}
Aviles Verdera, J., Story, L., Hall, M., Finck, T., Egloff, A., Seed, P. T., Malik, S. J., Rutherford, M. A., Hajnal, J. V., Tomi-Tricot, R., et al. (2023).
\newblock Reliability and feasibility of low-field-strength fetal MRI at 0.55 T during pregnancy.
\newblock \textit{Radiology}.
\newblock 309(1), e223050.
\newblock \url{https://doi.org/10.1148/radiol.23001478}.

\bibitem{payette2023automated}
Payette, K., Uus, A., Aviles Verdera, J., Avena Zampieri, C., Hall, M., Story, L., Deprez, M., Rutherford, M. A., Hajnal, J. V., Ourselin, S., et al. (2023).
\newblock An automated pipeline for quantitative T2* fetal body MRI and segmentation at low field.
\newblock \textit{International Conference on Medical Image Computing and Computer-Assisted Intervention}.
\newblock 358--367.
\newblock \url{https://doi.org/10.1007/978-3-030-70103-7_34}.

\bibitem{Isensee2021}
Isensee, F., Jaeger, P. F., Kohl, S. A. A., Petersen, J., Maier-Hein, K. H. (2021).
\newblock \textit{nnU-Net}: a self-configuring method for deep learning-based biomedical image segmentation.
\newblock \textit{Nature Methods}.
\newblock 18(2), 203--211.
\newblock \url{https://doi.org/10.1038/s41592-020-01008-z}.

\bibitem{PerezGarcia2021}
Pérez-García, F., Sparks, R., Ourselin, S. (2021).
\newblock TorchIO: A Python library for efficient loading, preprocessing, augmentation and patch-based sampling of medical images in deep learning.
\newblock \textit{Computer Methods and Programs in Biomedicine}.
\newblock 208, 106236.
\newblock \url{https://doi.org/10.1016/j.cmpb.2021.106236}.

\bibitem{Uus2021UNET}
Uus, A., Grigorescu, I., van Poppel, M., Hughes, E., Steinweg, J., Roberts, T., Lloyd, D., Pushparajah, K., Deprez, M. (2021).
\newblock 3D UNet with GAN discriminator for robust localisation of the fetal brain and trunk in MRI with partial coverage of the fetal body.
\newblock \textit{bioRxiv}.
\newblock \url{https://doi.org/10.1101/2021.06.23.449574}.
\newblock \url{https://www.biorxiv.org/content/early/2021/06/24/2021.06.23.449574}.

\bibitem{Neves2024}
Neves Silva, S., McElroy, S., Aviles Verdera, J., Colford, K., St Clair, K., Tomi-Tricot, R., Uus, A., Ozenne, V., Hall, M., Story, L., et al. (2024).
\newblock Fully automated planning for anatomical fetal brain MRI on 0.55 T.
\newblock \textit{Magnetic Resonance in Medicine}.
\newblock \url{https://onlinelibrary.wiley.com/doi/10.1002/mrm.29000}.

\end{thebibliography}
\end{document}